\documentclass[useAMS,usenatbib]{mn2e} \usepackage{graphicx}
\title[Self-consistent computation of $\gamma$-ray spectra due to
proton-proton interactions  in black hole systems] {Self-consistent
computation of $\gamma$-ray spectra due to proton-proton
interactions  in black hole systems}  \author[S. Bhattacharyya,
N. Bhatt and R. Misra]{S. Bhattacharyya$^1$, N. Bhatt$^1$ and
R. Misra$^2$\thanks{E-mail: rmisra@iucaa.ernet.in}\\
$^1$Astrophysical Sciences Division, Bhabha Atomic Research Centre,
Mumbai-$400085$, India \\ $^2$Inter-University Center for Astronomy
and Astrophysics, Post Bag 4, Ganeshkhind, Pune-$411007$, India}
\begin{document}

\date{Accepted ........ Received .........; in original form ........}

\pagerange{\pageref{firstpage}--\pageref{lastpage}} \pubyear{2006}

\maketitle

\label{firstpage}

\begin{abstract}

In the inner regions of an accretion disk around a black hole,
relativistic protons can  interact with ambient matter to produce
electrons, positrons and $\gamma$-rays.  The resultant steady state
electron and positron  particle distributions are self-consistently
computed taking into account Coulomb and Compton cooling, $e^-e^+$
pair production (due to $\gamma-\gamma$ annihilation) and pair
annihilation. While earlier works used the diffusion approximation to
obtain the particle distributions, here  we solve a more general
integro-differential equation that correctly takes into account the
large change in particle energy that occur when the leptons Compton
scatter off hard X-rays.  Thus this formalism can also be applied to
the hard state of black hole systems, where the dominant ambient
photons are hard X-rays.  The corresponding photon energy spectrum is
calculated and compared with broadband data of black hole binaries in
different spectral states.  The results indicate that the $\gamma$-ray
spectra ($E > 0.8$ MeV) of both the soft and hard spectral states and
the entire hard X-ray/$\gamma$-ray spectrum of the ultra-soft state,
could be due to $p-p$ interactions.  These results are consistent with
the hypothesis that there always exists in these systems a
$\gamma$-ray spectral component due to $p-p$ interactions which can contribute
between $0.5$ to $10$\% of the total bolometric luminosty. The model predicts
that {\it GLAST} would be able to detect black hole binaries and provide
evidence for the presence of non-thermal protons which in turn would
give insight into the energy dissipation process and jet formation in
these systems.

\end{abstract}


\section{Introduction}
Black hole X-ray binaries are generally observed to be in two distinct
states.    These states, which are named hard and soft, differ in
their luminosity and spectral shapes.  In the hard state, which is in
general less luminous, the spectrum of the system can be described as
a hard power-law with a spectral index $\Gamma \sim 1.7$ and a cutoff
around $100$ keV. This spectrum can be modeled  as thermal
Comptonization of soft photons by a plasma having temperature $T \sim
50$  keV \citep[e.g.][]{Gier97}. In contrast, the spectrum during the
soft state  consists of a blackbody-like component with $kT \sim 1$
keV, which typically dominates the luminosity and is generally
considered to be thermal emission from an optically thick accretion
disk. Apart from this soft (or disk)  component, a hard X-ray
power-law tail,  with  photon index $\Gamma \sim 2.5$ and no
detectable cutoff up to  $\sim 8$ MeV \citep{mcc02}, is also observed.

There have been several interpretations of the high energy ($E > 200$
keV) emission from black hole systems. It may arise from a photon
starved inner most region of a disk which cools due to bremsstrahlung
self-Comptonization \citep{Mel93} or as  emission due to $\pi^0$
decay, which are created by  proton-proton interaction in a hot proton
gas, $T > 10^{11}$ K \citep{Kol79,Jou94}. While these possibilities
maybe viable, they are based on assumptions of the geometry and
physical properties of the system, which are not directly and
independently verifiable, like the presence of a very hot proton gas
or a photon starved region. Another interpretation is that the spectra
arises due to Comptonization  of photons by the bulk motion of matter
falling into the black hole \citep{Lau99}.  However, \cite{Nie06}
have argued   that the non-detection of spectral breaks at $E < 500$
keV is contrary to this model's prediction.  A detailed radiative
model, which has been used to fit good quality broad band data of
black hole systems, is the hybrid model which is inscribed in a
spectral fitting code called EQPAIR \citep{cop98,Gier99}.  In the
framework of this model, this high energy component arises from   a
plasma  consisting of both thermal and non-thermal electrons  that
Comptonize external soft photons. Detailed spectral modeling of the
observed soft state spectra demands the coexistence  of thermal and
non-thermal electrons in the emission region and hence the steady
state non-thermal electron distribution is computed by assuming that
there is an injection of non-thermal  particles into the system where
they  cool by Comptonization and Coulomb interactions.

The origin of these non-thermal particles is uncertain. A possible
site may be a corona on top of a cold disk which is heated by magnetic
field reconnections \citep{har93, pout99}, but the details of the
process are largely unknown. The acceleration process has to be highly
efficient to produce non-thermal electrons in an environment where
electrons cool rapidly by inverse Comptonization. If this acceleration
process is  mass-independent then protons are also expected  to be
accelerated to relativistic energies, for example by scattering off
magnetic "kinks" in  a Keplerian accretion disk \citep{sub96}. Some
of these non-thermal protons may escape from the system and contribute
to the jet formation \citep{sub99}. This is particularly interesting
since there are some evidence that the X-ray producing region may be
same as the base of the extended jet \citep{Mar05}.  Hence the
detection of these non-thermal protons will provide valuable clues to
the nature of black hole systems.

Non-thermal protons would interact with the ambient thermal protons
and produce electron-positron  pairs, which would Comptonize photons
to high energies. These high energy photons would produce further
pairs by $\gamma-\gamma$ interaction and a pair cascade would
ensue. Pair cascades initiated by the injection of pairs (or equivalently high
energy non-thermal electrons)  have been extensively
examined \citep[e.g.][]{zdzlight,sven}. The effect of  $p-p$
interactions and  the resultant spectra have also been computed and
studied \citep[e.g.][]{sss,E80,E83,maha,sera,zdz86}.  These works, in
general, did not consider the presence of copious photons or have
assumed that the ambient photons are in the UV range, a scenario
relevant to AGN and under-luminous black hole systems.  However, for
black hole binaries, the system is dominated by either soft or hard
X-rays depending on the spectral state. One of the important radiative
interaction that would occur in such an environment is the inverse
Compton scattering of X-rays by pairs with Lorentz factor $\gamma
\approx 200$ that are produced by the $p-p$ interaction.  The standard
methods to compute the inverse Compton spectra, assume that the
interaction in the rest frame of the electron takes place in the
non-relativistic Thompson limit, which is true only if electron
Lorentz factor  times the photon energy $\gamma \epsilon$ $ \ll 
m_ec^2$.  This assumption is violated if the ambient photon energy is
$ \le m_ec^2/\gamma \approx 2$ keV.  Thus, when \citet{bhat03}
considered $p-p$ interaction in the presence of blackbody  photons
having temperature $T \approx 1$ keV, they used  a general formalism
to describe the inverse Compton process given by \citet{blu70}. They
found that for such a situation, which is relevant to the soft state
of black hole binaries, the effect of non-thermal protons is to
produce a broad feature around $1-50$ MeV. Using the observed OSSE
data for GRS 1915+105, they could constrain the fraction of
non-thermal protons in the system to be $< 5\%$.  Although
\citet{bhat03}, computed the change in energy of the photon (and hence
the change in energy of the lepton) appropriately in  the
Klein-Nishina regime,  they used, for simplicity, a diffusion equation
to describe the kinetic evolution of the pairs, which intrinsically
assumes that the change in energy of the particle per scattering is
small. This assumption breaks down when the ambient photon energy is
in X-rays, especially when there are a copious amount of ambient hard
X-rays.

In this work, we extend the formalism developed by \citet{bhat03}, by
solving an integro-differential equation for the pair kinetic
evolution which correctly takes into account the large energy changes
that a lepton undergoes upon scattering with an X-ray photon. This not
only allows for a more accurate estimation of the emergent spectra for
systems in the soft state, but enables the scheme to be applied to the
hard state also.

In \S $2$ we describe the model and the assumptions made to compute
the steady state non-thermal  electron/positron distributions and the
resultant photon spectra. In \S $3$ some general results of the
computation are presented along with comparison with observations of
black hole systems Cyg X-1  and GRS 1915+105 in different spectral
states.  The main results of the work are summarized and discussed in
\S $4$.

\section{Pair distribution and radiation spectra computation}

We consider a uniform sphere of non-relativistic thermal plasma with
number density $n_T$ and radius $R$, in the presence of an external
copious photon source. It is convenient to parameterize the luminosity
of the external photons  $L_{ph}$, in terms of the compactness
parameter $l_{ph} \equiv L_{ph} \sigma_T/(R m_ec^3)$, where $\sigma_T$
is the Thomson cross-section. The spectral shape of the ambient
photons is taken either to be a Wien peak or an exponentially cutoff
power-law depending on the spectral state being considered. In this
system, we assume that there is a power-law distribution  of
non-thermal protons with index $\alpha$, which would lead to
proton-proton collisions i.e.  the density of non-thermal protons is
given by
\begin{equation}
n_{NT} (\gamma) d\gamma = n_o (\gamma-1)^{-\alpha} d \gamma
\end{equation}
 The  normalization $n_o$ of this distribution is  characterized by
the compactness parameter $l_{p-p} \equiv L_{p-p} \sigma_T/(R
m_ec^3)$, where $L_{p-p}$ is the total power in electron, positrons
and $\gamma$-rays which would be produced in the proton-proton
interactions.

The steady state positron density $N_{+}(\gamma)$ is determined by
solving the  integro-differential equation
\begin{eqnarray}
\frac{\partial}{\partial \gamma}(\dot \gamma_C N_{+}(\gamma)) +
N_{+}(\gamma) \int^{\gamma}_{1} d\gamma^{\prime} P(\gamma,
\gamma^{\prime})- \nonumber \\   \int^{\infty}_{\gamma}
d\gamma^{\prime} P(\gamma^{\prime}, \gamma)N_{+}(\gamma^{\prime}) +
\dot N_{+}(\gamma) = Q_{+, pp}(\gamma) + Q_{+, \gamma \gamma}(\gamma)
\label{IDpos}
\end{eqnarray}
while the corresponding electron density $N_{-}(\gamma)$ is obtained
from
\begin{eqnarray}
\frac{\partial}{\partial \gamma}(\dot \gamma_C N_{-}(\gamma)) +
N_{-}(\gamma) \int^{\gamma}_{1} d\gamma^{\prime} P(\gamma,
\gamma^{\prime}) - \nonumber \\ \int^{\infty}_{\gamma}
d\gamma^{\prime} P(\gamma^{\prime}, \gamma)N_{-}(\gamma^{\prime}) =
Q_{-, pp}(\gamma) + Q_{-, \gamma \gamma}(\gamma)\label{IDele}
\end{eqnarray}

Here, $P(\gamma,\gamma-e)dedt$ is the probability that a
positron/electron with Lorentz factor  $\gamma$ will suffer a
collision and its Lorentz factor changes by an amount between $e$ and
$e+de$  in time $dt$, where $e \equiv h\nu/m_ec^2$ is the normalized
photon energy.  $Q_{\pm,\gamma \gamma}$ and $Q_{\pm, p p}$  are the
creation rates of pairs due to $p-p$ interactions and photon-photon
production respectively,   $\dot N_+(\gamma)$ is the annihilation
rates of positrons with the thermal background electrons and $\dot
\gamma_C$ is the Coulomb cooling rate.  This formalism to obtain the
particle distribution follows from the results described in
\citet{blu70} (and references therein), where the exact expression for
$P(\gamma, \gamma-e)dedt$ has been derived.  For small changes in
particle energy, eqs. (\ref{IDpos}) and (\ref{IDele}) can be reduced
to diffusion equations \citep{blu70} i.e. the integral terms can be
reduced to  ${\partial \over \partial \gamma} [\dot \gamma_{IC}
N_{\pm} (\gamma)]$ where $\dot \gamma_{IC}$ is the inverse Compton
cooling rate .  It was these diffusion equations that were used by
\citet{bhat03} as a simplifying assumption.

Following \citet{E83}, $e^+e^-$ production rate due to $p-p$ process
is given by
\begin{equation}
Q_{\pm,pp}(\gamma)=n_T c
\int^{\gamma_{p,h}(\gamma)}_{\gamma_{p,l}(\gamma)}
\frac{\sigma_{\pi^{\pm}}(\gamma_p)\beta_p n_{NT}(\gamma_p)}{\Big[(\bar
\gamma_{\star}-1)(2 \gamma^{3/4}_{pk} + \gamma_{pk}^{3/2})\Big]^{1/2}}
d\gamma_p
\end{equation}
where $\beta_p \equiv (1-1/\gamma_p^2)^{1/2}$ and $\gamma_{pk} \equiv
1-\gamma_p$. The denominator of the integrand represents the
appropriate energy distribution of the leptons produced, while the
limits impose the allowed energy range and are given by
\begin{equation}
\gamma_{p,h}(\gamma)=1+[\bar \gamma_{\star}\gamma + (\bar
\gamma_{\star}^2-1)^{1/2}(\gamma^2-1)^{1/2}-1]^{4/3}
\end{equation}
and
\begin{equation}
\gamma_{p,l}(\gamma)=1+[\bar \gamma_{\star}\gamma - (\bar
\gamma_{\star}^2-1)^{1/2}(\gamma^2-1)^{1/2}-1]^{4/3}
\end{equation}
where $\bar \gamma_{\star}\equiv 70$. The approximate but analytical
cross-section ($\sigma_{\pi^{\pm}}$) is tabulated for different energy
ranges by \citet{E83}.

 Pair production rate from  photon-photon interactions is approximated
 to be
\begin{equation}
Q_{\pm,\gamma\gamma}(\gamma)=c \int n_{\gamma}(2\gamma_e-e)
n_{\gamma}(e)\sigma_{\gamma \gamma}(e,2\gamma-e)de
\end{equation}
where it has been assumed that for two photons annihilating with
energies $e$ and $e^\prime$, the resultant  Lorentz factor is
$\sim(e+e^\prime)/2$. The approximate form for the  cross section
$\sigma_{\gamma \gamma} (e,e^\prime)$ is given by \citet{CB90}.  The
positrons primarily annihilate with the background  thermal electrons
at a rate given by
\begin{equation}
\dot N_+(\gamma)=N_+(\gamma)n_T \sigma_{e^+e^-}(1,\gamma)c
\end{equation}
where the approximate form of the cross section is given by
\citet{CB90}.

It is worthwhile to note that, while the numerical computation of the
integro-differential equations (eqs. \ref{IDpos} and \ref{IDele}) are
relatively straight forward, there are a few minor but tricky points
that have to be taken into consideration. In particular, the number
conservation of photons for the inverse Compton process has to be
strictly satisfied in the numerical computation, in order to obtain
the correct and stable particle density solution.

The equilibrium photon density inside the sphere is a solution of
\begin{equation}
Q_{\gamma,IC}+Q_{\gamma,pp}+Q_{\gamma,e^+e^-}=n_{\gamma}(e)\Big[R_{\gamma
\gamma}+\frac{c}{R[1+\tau_{KN}(e)]}\Big]
\label{Phoden}
\end{equation}
where $R_{\gamma \gamma}(e)$ is the rate of photon
annihilation and  $\tau_{KN}$ is the Klein-Nishina optical depth.
$Q_{\gamma,IC}$, $Q_{\gamma,pp}$ and
$Q_{\gamma,e^+e^-}$ are the photon production rates due to inverse
Compton, $p-p$ interaction and pair annihilation respectively. The
photon production due to $p-p$ reaction is given by
\begin{equation}
Q_{\gamma,pp}(e)=\frac{m_e}{m_{\pi}}n_Tc\int_{\gamma_{p,l}(e)}^{\infty}
\frac{\sigma_{\pi^o}(\gamma_p)\beta_p
n_{NT}(\gamma_p)d\gamma_p}{(\gamma_{pk}^{1/2}+2\gamma_{pk}^{1/4})^{1/2}}
\end{equation}

For simplicity, the  scattering of high energy photons with the
thermal particles are neglected here and the implications of this
assumption are discussed in the last section.  The interactions of
photons with high energy protons, which requires a much larger
threshold proton energy ($\gamma_p > 300$), have also been
neglected. As noted by \citet{bhat03}, such interactions are not
important when $\alpha > 2$ and depend on the unknown high energy
cutoff of the proton acceleration process.

Equations (\ref{IDpos}), (\ref{IDele}) and (\ref{Phoden}) are solved
self-consistently to obtain electron and positron distributions as
well as the radiative flux. The output spectrum depends on the
spectral shape of the external photons and three other parameters: the
Thomson optical depth $\tau$, photon compactness $l_{ph}$ and the
ratio  of proton-proton compactness $l_{p-p}$, to that of the external
photon, $\beta \equiv l_{p-p}/l_{ph}$.  The results are insensitive to
the non-thermal proton index $\alpha$ and apart for an overall
normalization factor, to the size $R$ of the system. From the
definition of compactness ($L \propto l R$), it follows that for
a fixed value of $l$, the normalization of the photon spectrum
will be proportional to $R$. The spectral shape primarily depends
on the inverse Compton and pair production processes, which are
characterized by $\tau_{IC} \sim n_{\gamma,s} \sigma_T R$ and 
$\tau_{\gamma\gamma} \sim n_{\gamma}  \sigma_T R$. Since the
soft photon density $n_{\gamma,s} \propto L_{ph}/R^2 \propto l_{ph}/R$
and similarly, the high energy photon density, $n_{\gamma} \propto l_{p-p}/R$,
$\tau_{IC}$ and $\tau_{\gamma\gamma}$ are independent of $R$ for specified
compactness. Thus, parameterizing the luminosities in terms of compactness
renders the shape of the spectrum to be nearly independent of the size of
the system.

\section{Results}

\subsection{Soft State}

\begin{figure}
\includegraphics[width=8cm,height=8cm,angle=-90]{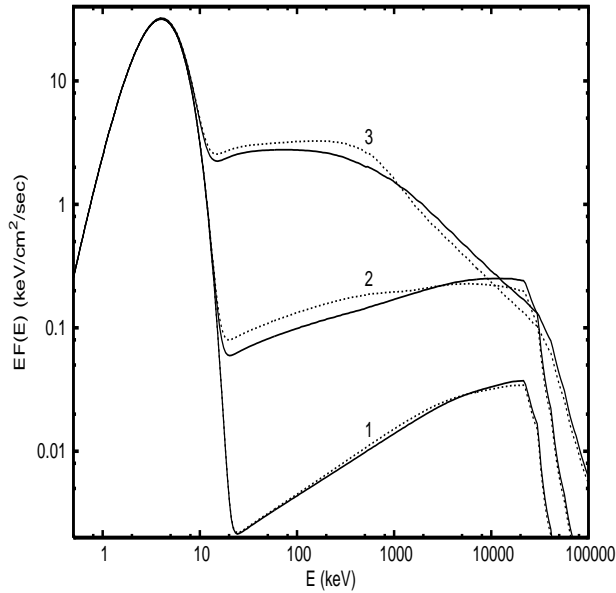}
\caption{The radiated spectra  for parameters typical of the soft
state of black hole binaries,which are $l_{ph} = 250$, $kT_{ph}=1$
keV,  $\tau=2.5$ and $\alpha=2.5$.  The three curves labeled as $1$,
$2$ and $3$  are for compactness ratio $\beta \equiv l_{p-p}/l_{ph} =
0.001$, $0.01$  and $0.1$ respectively. The solid lines represent
spectra obtained using the integro-differential approach to solve for
the particle energy distribution.  The dotted lines represent spectra
obtained using the diffusion approximation as described in
\citet{bhat03}. }
\label{Soft_gen_spec_a}
\end{figure}

\begin{figure}
\includegraphics[width=8cm,height=8cm,angle=-90]{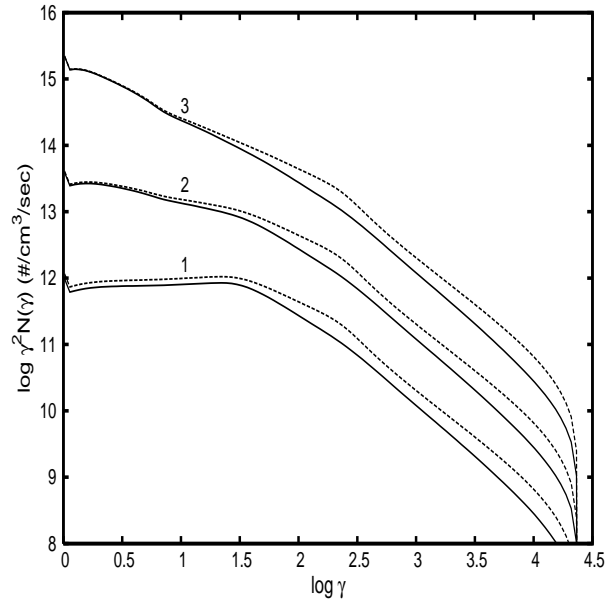}
\caption{The particle distributions corresponding to the spectra shown
in Figure (\ref{Soft_gen_spec_a}). The dotted (solid) lines represent
the positron (electron) distributions.}
\label{Soft_gen_spec_b}
\end{figure}

The computed radiated spectra due to $p-p$ interactions, corresponding
to parameters typical of the soft state of black hole binaries, are
shown in Figure \ref{Soft_gen_spec_a}.  Since the motivation of this
work is to investigate the possibility that the $\gamma$-ray part of
the spectrum is due to $p-p$ interactions, a detailed fit to the broad
band data, especially the low energy part of the spectrum, is not
being attempted. Instead, the dominant soft component is being
represented by a a Wien peak like spectrum at $kT_{ph} = 1$ keV and
high compactness,  $l_{ph} = 250$. The corresponding luminosity is
\begin{equation}
L \approx  10^{38} \; \hbox {ergs s}^{-1}\;\;\; \Big
(\frac{l_{ph}}{250}\Big ) \;\; \Big (\frac{R}{10^7 \hbox {cm}}\Big)
\end{equation}
where $R \approx 10^7$ cm is the size of the system.  The  optical
depth of the thermal electrons is of order unity  and is taken to be
$\tau = 2.5$ in Figure \ref{Soft_gen_spec_a}.  The spectra plotted in
the figure, correspond to three different values of the compactness
ratio $\beta \equiv l_{p-p}/l_{ph}  = 0.001$, $0.01$ and $0.1$, which
represent increasing efficiency of the acceleration process to produce
non-thermal protons. At low $\beta$ values the spectra are hard with a
break around $\sim 80$ MeV. This cutoff is due to the interaction of
higher energy photons  with the ambient soft photons of energy $\sim
3$ keV to generate pairs.  The hard spectra are due to inverse
Comptonization of photons by a steady state pair distribution
$N_{\pm} (\gamma) \propto \gamma^{-2}$ (Figure
\ref{Soft_gen_spec_b}). Such a distribution is expected when pairs are
injected at a large energy ($\gamma \approx 200$) and are cooled by
inverse Comptonization. Essentially in the diffusion approximation,
the cooling rate $\dot \gamma_{IC} \propto \gamma^2$, which leads to
$N (\gamma) \propto \gamma^{-2}$ when pair creation due to
photon-photon interaction can be neglected. For higher compactness
ratio, $\beta = 0.01$, pair creation due to photon-photon interaction
affects the particle distribution at low $\gamma$ (Figure
\ref{Soft_gen_spec_b}), which leads to a softer output spectra
(Figure \ref{Soft_gen_spec_a}). For still larger values of $\beta =
0.1$, apart from the change in particle distribution due to
photon-photon pair production, another break at around $511$ keV
appears in the radiated spectrum, since the density of high energy
photons is large enough for pairs to be produced by self interaction.

Figure \ref{Soft_gen_spec_a} also shows for comparison, the computed spectra 
obtained using
the diffusion approximation to solve for the particle distribution
as described in \citet{bhat03}. While, quantitative differences can be seen
when the compactness ratio is large (spectra marked 2 and 3), the
qualitative features of the spectra are similar.

\begin{figure}
\includegraphics[width=8cm,height=8cm,angle=-90]{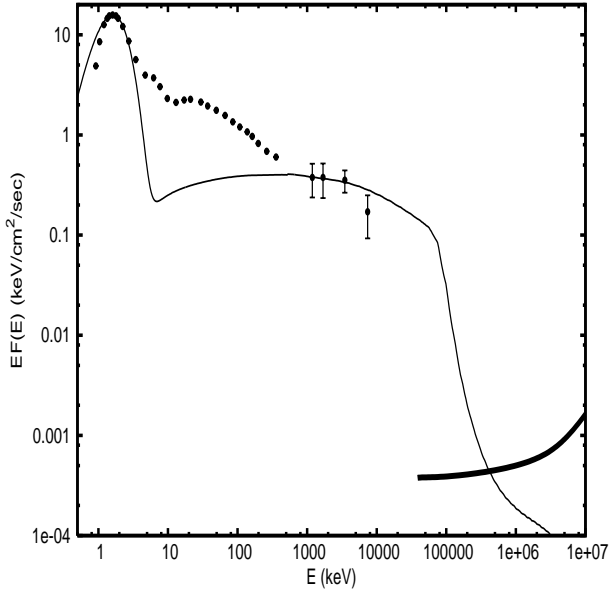}
\caption{Broad band soft state data of Cyg X-1 reproduced from
\citet{mcc02}. Solid line is the computed spectrum for $\beta = 0.046$,
$l_{ph} = 125$, $kT_{ph}=0.4$ keV, $\tau=2.5$ and $\alpha=2.5$. The
thick broad line shows the estimated {\it GLAST} sensitivity
(http://www-glast.slac.stanford.edu/software/IS/glast$\_$lati$\_$performance.htm).}
\label{Cyg_soft}
\end{figure}

The high energy (5--50) soft state spectra are characterized by a
steep power-law of photon index $\Gamma \approx 2.5$, and hence cannot
be explained as emission from $p-p$ interactions, since in this energy
range the expected spectral index is $\Gamma < 2$ (Figure
\ref{Soft_gen_spec_a}). Indeed, the soft state spectra can be
described by the hybrid model, where non-thermal electrons are
injected into a plasma. In Figure \ref{Cyg_soft} we reproduce the
unfolded data points based on such a model (the  ``EQPAIR'' model) as
fitted by \citet{mcc02}. These data obtained from instruments on board
the {\it BeppoSAX} and {\it CGRO} satellites cover a wide energy range
from $\approx 0.5$ keV to $\approx 8$ MeV. As shown by \citet{mcc02},
the spectrum at energies $< 1$ MeV can be represented as an extension
of the non-thermal spectra at hard X-rays. On the other hand, the
$\gamma$-ray (i.e. the 1--8 MeV) spectrum, maybe due to an
additional component, especially since there are uncertainties about
the relative normalization of the different instruments used to make
the composite spectrum. A second component interpretation is supported
by the non-detection by COMPTEL of any flux in the energy range from
$750$ keV to $1$ MeV, indicating a possible spectral break in that
region, although the low statistics of the data does not allow for a
definite conclusion.  Thus, we attempt to fit the $\gamma$-ray part of
the spectrum, as emission from $p-p$ interactions, using parameters
that are consistent with those obtained by \citet{mcc02}. The required
compactness ratio $\beta = l_{p-p}/l_{ph}$ $\approx 0.046$, can be
compared with $l_{nt}/l_{ph} = 0.11$ obtained by the EQPAIR fit. This
indicates that if the $\gamma$-ray spectrum is due to $p-p$
collisions, the energetics of the proton acceleration that produces
non-thermal protons is comparable to the electron one.  The factor of
two difference, between the two ratios, could be either due to the
neutrino emission, which has not been added to $l_{p-p}$ or to the
uncertainties involved in the theoretical modeling and fitting of the
data. If the $\gamma$ ray spectrum is not due to this process, the
above estimate serves as an upper limit on the  energetics of the
non-thermal acceleration of the protons. 
For the parameters used to compute the p-p component, the
ratio  of the non-thermal proton density to
the thermal one turns out to be $\approx 0.03$, 
which is consistent with that estimated earlier by \citet{bhat03}.

\citet{zdz04} have estimated the sensitivity of a $4 \times 10^5$ s
{\it GLAST} observation to be $\approx 10^{-3}$ keV cm$^{-2}$ s$^{-1}$ at 10
GeV. Figure \ref{Cyg_soft} shows that the predicted spectrum at that
energy is a couple of order of magnitude higher than this limit and
would be easily detectable. The extension of the hybrid plasma model
fit to the data \citep{mcc02} to 10 GeV also predicts a similar
flux. However, as pointed out by \citet{zdz04}, the hybrid plasma
model assumes a low value of compactness $l_{ph} \approx 4$ and a high
value of the maximum Lorentz factor of the electrons, $\gamma_{max} =
10^4$. For a more realistic value of compactness or if electrons are
not accelerated to such high energies, there will be a sharp cutoff in
the hybrid plasma spectrum at $\approx 1$ GeV. Thus, a {\it GLAST}
observation of Cyg X-1 (or black hole binaries in general) during the
soft state, should be able to discern between these models.

\begin{figure}
\includegraphics[width=8cm,height=8cm,angle=-90]{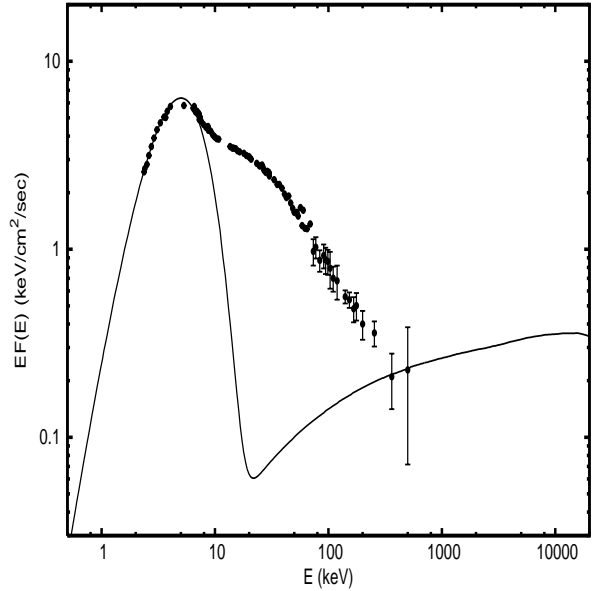}
\caption{Broad band Very-High state ($\chi$ class) data of GRS
1915+105 reproduced from \citet{zdz01}.  Solid line is computed
spectrum for $\beta = 0.075$,  $l_{ph} = 50$, $kT_{ph}=1.25$ keV,
$\tau=2.5$ and $\alpha=2.5$}
\label{GRSchi}
\end{figure}

Black hole systems exhibit a variation of the soft state spectrum,
which  is sometimes denoted as a separate state and is called Very
High or Intermediate state (VHS/IS). Here the hard X-ray power-law is
nearly equally luminous as the soft component. The black hole system
GRS 1915+105 displays a wide variety of spectral shape and its $\chi$
class behavior (as classified by \citet{Bel00})  is similar to the VHS
state of other black hole systems. \citet{zdz01} have fitted the broad
band data of this source obtained from instruments on board {\it RXTE}
and {\it CGRO}, by the hybrid model and the unfolded data based on
that  model is  reproduced in Figure \ref{GRSchi}. The absence of high
energy COMPTEL data does not allow for detailed fits, but a
comparison of the $600$ keV flux with the computed spectrum provides an
upper limit of the compactness ratio $\beta = 0.075$,  which again can
be compared to the ratio $l_{nt}/l_{ph} = 0.11$ required by the hybrid
model fit \citep{zdz01}.

\begin{figure}
\includegraphics[width=8cm,height=8cm,angle=-90]{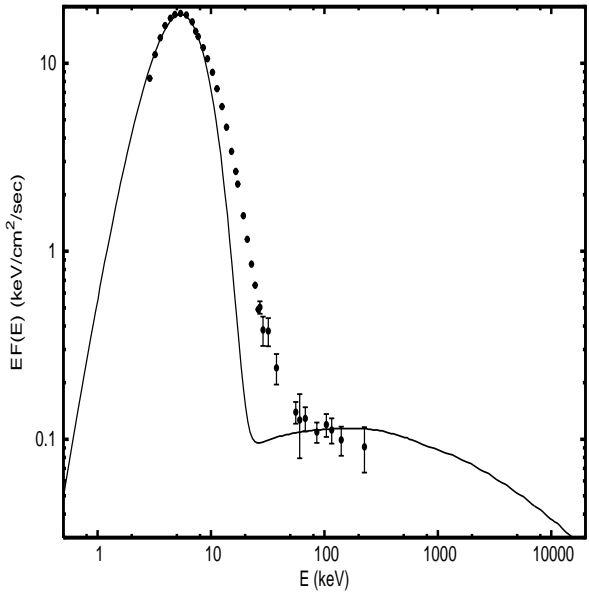}
\caption{Broad band Ultra-soft state ($\gamma$ class) data of GRS
1915+105 reproduced from \citet{zdz01} Solid line is computed spectrum
for  $\beta = 0.008$,  $l_{ph} = 555$, $kT_{ph}=1.35$ keV, $\tau=5.0$
and $\alpha=2.5$}
\label{GRSgamma}
\end{figure}

Another variation of the soft state spectra is the so called
ultra-soft state, where the spectrum is dominated by the soft
blackbody like emission, with a weak hard X-ray tail.  This state has
been observed in many black hole systems like GRS 1915+105, GX 339-4
and XTE J1550-564.  An example of this state is shown in Figure
\ref{GRSgamma}. Here the data is reproduced from \citet{zdz01}, and is
unfolded based on the EQPAIR model fit to observation of GRS 1915+105
in the $\gamma$ class using {\it RXTE} and {\it CGRO} satellites. A
remarkable observational feature is that for all these sources,  the
weak high energy tail is hard with a photon spectral index $\Gamma
\approx 2$, for which currently there is no theoretical explanation
\citep{zdz04}. For high compactness the spectra due to $p-p$
interaction, is expected to have such a hard spectral slope (Figure
\ref{Soft_gen_spec_a}). Indeed as shown in Figure \ref{GRSgamma}, the
hard X-ray spectra can be explained solely as emission from such a
process with a compactness ratio $\beta \approx 0.008$. Again, the
absence of data greater than $1$ MeV does not allow for a clear
identification. However, the data is consistent with the hypothesis,
that in the Ultra-soft state, the emission from accelerated
non-thermal electrons is absent and hence, only the hard spectra
arising from $p-p$ interactions is observed.

\subsection{Hard State}

\begin{figure}
\includegraphics[width=8cm,height=8cm,angle=-90]{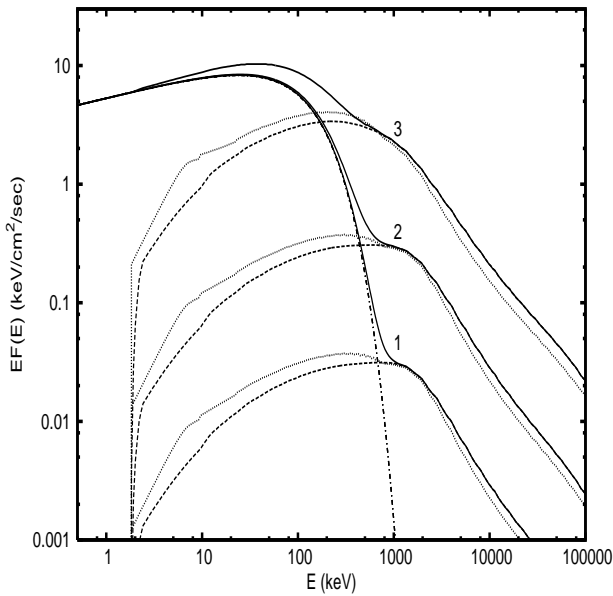}
\caption{The radiated spectra  for parameters typical of the hard
state of black hole binaries, which are  $l_{ph} = 250$, $E_c=100$
keV, $\tau=1.0$ and $\alpha=2.5$. The three curves labeled as $1$, $2$
and $3$ are for compactness ratio $\beta \equiv l_{p-p}/l_{ph} =
0.001$, $0.01$  and $0.1$ respectively. The solid lines represent the
total spectra, the dotted lines represent the spectral component
due to $p-p$ interactions and the dot dashed line represents the
thermal component.  The dashed line represent the spectral
components obtained using  the diffusion approximation as described in
\citet{bhat03}}.
\label{Hard_gen_spec_a}
\end{figure}

\begin{figure}
\includegraphics[width=8cm,height=8cm,angle=-90]{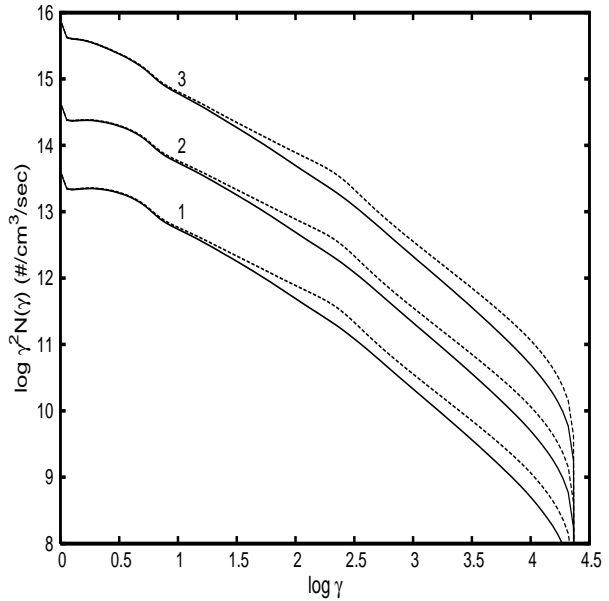}
\caption{The particle distributions corresponding to the spectra shown
in Figure (\ref{Hard_gen_spec_a}). The dotted (solid) lines represent
the positron (electron) distributions.}
\label{Hard_gen_spec_b}
\end{figure}

Figures \ref{Hard_gen_spec_a}  and  \ref{Hard_gen_spec_b} show the
computed spectra and particle densities for different values of
compactness ratio $\beta$, when the ambient photon density is similar
to that found in the hard state of black hole binaries. In this state,
the spectrum is dominated by a thermal Comptonized component, which is
approximated here as an exponentially cut off power law with photon
index $\Gamma = 1.7$ and cut off energy $E_c = 100$ keV. As the
average photon energy in the plasma is  higher than that of the soft
state, the cut-off in the photon spectrum due to pair production
appears at $\approx 3$ MeV instead of $\approx 80$ MeV. In the soft
state, the shape of  the computed spectra and particle density
depended on the compactness ratio $\beta$. This was due to increasing
self-interaction of the non-thermal photons to produce pairs as
compactness increased. However, in the hard state, since the number
density of ambient photons with energy $> 500$ keV is large, pair
production is dominated by interaction of non-thermal photons with the
background ambient ones even for large compactness. Thus, the shape of
the computed spectra is relatively invariant to $\beta$ as shown in
Figure \ref{Hard_gen_spec_a} and in particular the slope of the
expected spectrum at energies $>$ MeV is insensitive to the parameters
of the model.

Figure \ref{Hard_gen_spec_a} also shows for comparison, the computed spectral 
components due to $p-p$ interaction, obtained using
the diffusion approximation to solve for the particle distribution
as described in \citet{bhat03}. Similar to what was found for the
soft state (Figure \ref{Soft_gen_spec_a}), only quantitative differences can be seen
while the
qualitative features of the spectra are similar. This suggests that even
when the ambient photons have energy $ \sim 100$ keV, the diffusion approximation
can reveal the salient features of the high energy spectrum.

\begin{figure}
\includegraphics[width=8cm,height=8cm,angle=-90]{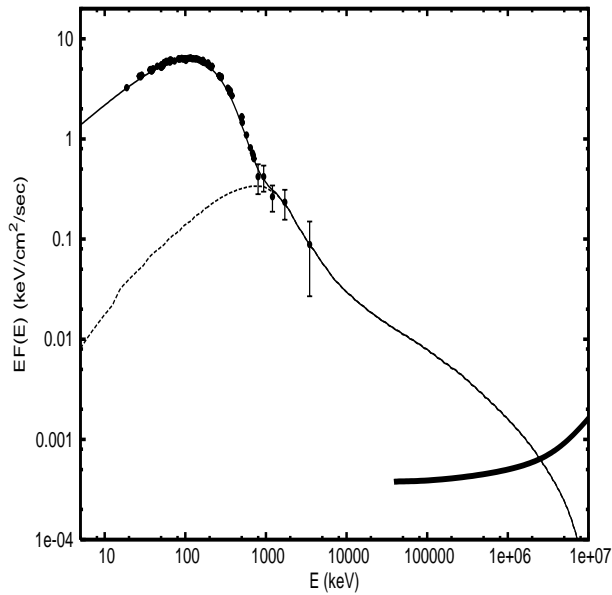}
\caption{Broad band hard state data of Cyg X-1 reproduced from
\citet{mcc02}. Solid line is computed spectrum for $\beta = 0.018$,
$l_{ph} = 110$, $E_{c}=120$ keV, $\tau=1.0$ and $\alpha=2.5$.  The
thick broad line shows the estimated {\it GLAST} sensitivity
(http://www-glast.slac.stanford.edu/software/IS/glast$\_$lati$\_$performance.htm).
The dotted line  represents the spectral component due to $p-p$
interactions.}
\label{Cyg_hard}
\end{figure}

The broadband hard state data of Cyg X-1, using {\it RXTE} and {\it
BeppoSAX} satellites, were analyzed and fitted using the EQPAIR model
by \citet{mcc02}. The unfolded data is reproduced in Figure
\ref{Cyg_hard}. Most of the spectra can be represented by thermal
Comptonization, but there is evidence for a different high energy
spectral component in the $0.7-8$ MeV range, which was explained as
emission from non-thermal electrons by \citet{mcc02} (see also
\citet{Ibr05}). Here we compare the data  with the computed spectra
due to $p-p$ interactions, using  parameters representative of the
hard state and show that the the MeV data, especially the spectral
slope in that energy range, match with the observations. The required
compactness ratio  $\beta = 0.018$ is similar to the values required
to fit  the soft state   data, but smaller than $l_{nt}/l_{h} = 0.08$
required by the EQPAIR fit \citep{mcc02}.  The overall computed
spectra fits even the data at energies $< 700$ keV remarkably well,
however, this could be a coincidence given that an exponentially
cutoff power-law is not a good approximation to a Comptonized thermal
spectrum and that reflection is not taken into account
here. Nevertheless, the fit in Figure \ref{Cyg_hard}, strongly
indicates that the emission from Cyg X-1 in the hard state can be
explained as thermal Comptonization and a component due to $p-p$
interactions.

The spectrum shown in Figure \ref{Cyg_hard}, predicts that {\it GLAST} should
be able to detect Cyg X-1 in the hard state also. Similar to the soft
state, if the high energy emission is due to non-thermal electrons
(like in the EQPAIR model fit by \citet{mcc02}) a detection by {\it GLAST}
would only be possible under extreme conditions like low compactness
and high maximum Lorentz factor of the non-thermal electrons
\citep{zdz04}. Thus, {\it GLAST} would be able to differentiate between
these models.

\section{Summary and Conclusion}

A self-consistent scheme, which computes the electron/positron and
radiation energy distribution inside a thermal plasma having
non-thermal protons, is developed. The non-thermal protons interact
via $p-p$ collisions to produce electron/positron pairs and
$\gamma$-rays.  These high energy pairs cool by inverse Comptonization
of ambient photons. The effect of subsequent pair production due to
photon-photon interactions and annihilation of positrons with ambient
electrons, is taken into account. Unlike previous works, the scheme
does not assume that the energy change of the leptons during inverse
Compton scattering is small. Hence it can be applied to black hole
binary systems, where the dominant ambient photons are X-rays.
 
Comparison of the computed spectra with the broad band
X-ray/$\gamma$-ray spectra of black hole binaries in different
spectral states reveal:
  
\noindent $\bullet$ The hard X-ray spectra (3--800 keV) of  the soft
and Very High state (VHS), are too steep to be explained as emission
due to $p-p$ interactions. However, the observed $\gamma$-ray spectra
(0.8--8 MeV), especially for the soft state,  could be due to such
interactions.  In this interpretation, an unknown acceleration process
energises both electron and protons, producing their non-thermal
distributions. While the electron non-thermal distribution gives rise
to the hard X-ray emission, the protons are the origin of the
$\gamma$-rays. The powers going into accelerating protons and
electrons are within a factor of two.
    
\noindent $\bullet$ For the ultra-soft state, the observed hard X-ray
spectra ( E $> 5$ keV) could be entirely due to $p-p$
interactions. This interpretation gives a natural explanation for the
similar hard X-ray spectral slope ($\Gamma \approx 2$) observed
whenever a black hole is in the ultra-soft state.

\noindent $\bullet$ For the hard state, the observed $\gamma$-ray
spectrum (0.5--8 MeV) can be explained as emission due to  $p-p$
interactions alone. The predicted steep spectral shape in this energy
range, is not sensitive to the model parameters, and matches well with
the observations.

\noindent $\bullet$ For both the soft and hard states, the model
predicts that for reasonable parameters, {\it GLAST} should be able to
detect black hole binaries. This in contrast to the situation when
only non-thermal electrons are present in the system, where very low
compactness and large maximum Lorentz factor of the electrons have to
be postulated, in order for {\it GLAST} to make a similar detection.

These results are consistent with the hypothesis that there always
exists a spectral component due to a non-thermal proton distribution
in black hole binaries.  This component peaks at $\gamma$-rays and 
can contribute between $0.5$ to $10$\% 
of the bolometric luminosity. In the
ultra-soft state, this component is visible for $E > 5$ keV. During
the soft and hard states, the emission is detectable only when $E >
0.8$ MeV, since  other spectral  components dominate at lower
energies. This hypothesis can be verified using future observations by
{\it GLAST}.

The scheme does not include scattering of  high energy photons  with
thermal electrons. Although the Klein-Nishina cross section decreases
with photon energy, such scatterings can be important  especially
when the Thompson optical depth is significantly greater than unity, a
case more pertaining to the soft state. In fact, the inability of the
present model to explain the hard X-ray emission during the soft
state, may be an artifact of this assumption. If that is true, then it
may be possible that only protons are accelerated in black hole
binaries and not electrons.  This is theoretically appealing given the
complexities of  accelerating electrons to high energies in a region
where  inverse Compton cooling is efficient. However, several
sophisticated modification of the present scheme have to be undertaken
before this speculation can be tested.

Evidence of non-thermal protons in black hole binaries, would shed
light on the energy dissipation process that occur in the  inner
regions of the accretion disk. These energetic protons could also be
the origin of the outflows/jets that are observed in many of these
systems. Such insights may finally lead to a comprehensive physical
picture of these enigmatic sources.

\section{Acknowledgments}
The authors would like to thank the referee, A. A. Zdziarski, for pointing
out a mistake in the initial draft and for making detailed recommendations
which has significantly improved the paper.

\bsp

\label{lastpage}


\begin{thebibliography}{}

\bibitem[\protect\citeauthoryear{Belloni et. al.}{2000}] {Bel00}
Belloni, T., Klein-Wolt, M., Mendez, M., van der Klis, M., van
Paradijs, J., 2000,  A\&A, 355, 271

\bibitem[\protect\citeauthoryear{Blumenthal \& Gould}{1970}]{blu70}
Blumenthal, G. R., Gould, R. J. 1970, Rev. Mod. Phys. 42, 237

\bibitem[\protect\citeauthoryear{Bhattacharyya et al.}{2003}]{bhat03}
Bhattacharyya, S., Bhatt, N., Misra, R., Kaul, C. L. 2003, ApJ, 595, 317

\bibitem[\protect\citeauthoryear{Coppi \&  Blandford} {1990}]{CB90}
Coppi, P. S., Blandford, R. D., 1990, MNRAS, 245, 453

\bibitem[\protect\citeauthoryear{Eilek}{1980}] {E80} Eilek, J. A.,
1980, ApJ, 236, 664

\bibitem[\protect\citeauthoryear{Eilek \& Kafatos}{1983}] {E83} Eilek,
J. A.,  Kafatos, M., 1983, ApJ, 271, 804

\bibitem[\protect\citeauthoryear{Gierli{\'n}ski et al.}{1999}]{Gier99}
Gierli{\'n}ski, M., Zdziarski A. A., Poutanen J., Coppi P. S.,
Ebisawa, K., Johnson W. N. 1999, MNRAS, 309, 496

\bibitem[\protect\citeauthoryear{Gierli{\'n}ski et al.}{1997}]{Gier97}
Gierli{\'n}ski, M., Zdziarski A. A., Done, C., Johnson W. N.,
Ebisawa K., Ueda, Y., Haardt, F. Phlips, B. F. 1997, MNRAS, 288, 958

\bibitem[\protect\citeauthoryear{Haardt \& Maraschi}{1993}]{har93}
Haardt, F. Maraschi, L. 1993, ApJ, 413, 507

\bibitem[\protect\citeauthoryear{Ibragimov et  al.}{2005}]{Ibr05}
Ibragimov A., Poutanen J., Gilfanov M.,  Zdziarski A. A., Shrader
C. R. 2005, MNRAS, 362, 1435

\bibitem[\protect\citeauthoryear{Jourdain \& Roques}{1994}]{Jou94}
Jourdain, E., Roques, J. P. 1994, ApJ, 426, L11

\bibitem[\protect\citeauthoryear{Kolykhalov \& Sunyaev}{1979}]{Kol79}
Kolykhalov, P.I., Sunyaev, R. 1979, Soviet. Astron., 23, 189

\bibitem[\protect\citeauthoryear{Laurent \& Titarchuk}{1999}]{Lau99}
Laurent, P.,  Titarchuk, L. 1999, ApJ, 511, 289

\bibitem[\protect\citeauthoryear{Lightman \& Zdziarski} {1987}]
{zdzlight}  Lightman, A. P.,  Zdziarski, A. A. 1987, ApJ, 319, 643

\bibitem[\protect\citeauthoryear{Mahadevan et al.}{1997}]{maha}
Mahadevan, R. Narayan, R.,  Krolik, J. 1997, ApJ, 486, 268

\bibitem[\protect\citeauthoryear{Markoff et al.}{1999}]{sera}
Markoff, S., Melia, F., Sarcevic, I. 1999, ApJ, 522, 870

\bibitem[\protect\citeauthoryear{Markoff et al.}{2005}]{Mar05}
Markoff, S., Nowak, M. A., Wilms, J.  2005, ApJ, 635, 1203

\bibitem[\protect\citeauthoryear{McConnell et al.}{2002}]{mcc02}
McConnell, M. L., et al. 2002,	 ApJ, 572, 984

\bibitem[\protect\citeauthoryear{Melia \& Misra}{1993}]{Mel93}
Melia, F.,  Misra, R. 1993, ApJ, 411, 797

\bibitem[\protect\citeauthoryear{Nied{\'z}wiecki \&
Zdziarski}{2006}]{Nie06} Nied{\'z}wiecki A., Zdziarski A. A.  2006,
MNRAS, 365, 606

\bibitem[\protect\citeauthoryear{Poutanen \& Coppi}{1998}]{cop98}
Poutanen, J., Coppi, P. S. 1998, Phys. Scr., T77, 57

\bibitem[\protect\citeauthoryear{Poutanen \& Fabian}{1999}]{pout99}
Poutanen, J., Fabian, A. C. 1999, MNRAS, 306, L31

\bibitem[\protect\citeauthoryear{Stern et al.}{1992}]{sss}  Stern,
B.E.,  Sikora, M. \& Svensson R., 1992, Testing the AGN paradigm,
Proceedings of the 2nd Annual Topical Astrophysics Conference, p. 313.

\bibitem[\protect\citeauthoryear{Subramanian, Becker \&
Kafatos}{1996}]{sub96} Subramanian, P., Becker, P.A., Kafatos,
M. 1996, ApJ, 469, 784

\bibitem[\protect\citeauthoryear{Subramanian et al.}{1999}] {sub99}
Subramanian, P., Becker, P. A., Kazanas, D. 1999, ApJ, 523, 203

\bibitem[\protect\citeauthoryear{Svensson}{1987}] {sven}  Svensson,
R., 1987, MNRAS, 227, 403

\bibitem[\protect\citeauthoryear{Zdziarski}{1986}]{zdz86} Zdziarski,
A. A. 1986, ApJ, 305, 45

\bibitem[\protect\citeauthoryear{Zdziarski et al.}{2001}]{zdz01}
Zdziarski, A. A. Grove, J. E., Poutanen, J., Rao, A. R., Vadawale, S. V.
2001,	ApJ, 554, L45

\bibitem[\protect\citeauthoryear{Zdziarski \&
Gierli{\'n}ski}{2004}]{zdz04} Zdziarski, A. A.\& Gierli{\'n}ski, M. 2004,
Prog. Theor. Phys. Suppl., 155, 99

\end{thebibliography}
\end{document}